\documentclass[
  ,draft            
  ]
  {aipproc}

\layoutstyle{8x11double}

\begin{document}

\title{Neutron Star Crustal Mass Fractions}


\author{Kelsey L. Hoffman\footnote{kelsey@phas.ubc.ca} }{
  address={Department of Physics and Astronomy, University of British
    Columbia, 6224 Agricultural Road, Vancouver, British Columbia V6T
    1Z1, Canada}
}

\author{Jeremy S. Heyl\footnote{Canada Research Chair}
}{
  address={Department of Physics and Astronomy, University of British
    Columbia, 6224 Agricultural Road, Vancouver, British Columbia V6T
    1Z1, Canada}
}


\begin{abstract}
  We are investigating mass fractions on the crust of a neutron star
  which would remain after one year of cooling. We use cooling curves
  corresponding with various densities, or depths, of the neutron star
  just after its formation. We assume the modified Urca process
  dominates the energy budget of the outer layers of the star in order to
  calculate the temperature of the neutron star as a function of
  time. Using a nuclear reaction network up to technetium, we
  calculate how the distribution of nuclei quenches at various depths
  of the neutron star crust. The initial results indicate that
  $^{28}$Si is the lightest isotope to be optically thick on the
  surface after one year of cooling. 
\end{abstract}


\maketitle


\section{Introduction}

The observed  emission from a neutron star passes through a 
crustal layer of the neutron star. In order to fully interpret the observed
emission from a neutron star, we need to understand what comprises this
crustal layer of the neutron star \cite{Hern84b,heylmnras01}. Here, the mass
fractions on the surface are calculated for what would exist on the
crust after one year of cooling of the neutron star. These mass
fractions are calculated using a 489 isotope reaction network which
burns up to technetium written by F.X. Timmes \cite{timmesapjs99} and
is available on Timmes's webpage
\footnote{\url{http://www.cococubed.com/code_pages/burn.shtml}}.

\section{Cooling Curve and Input Temperature}

\begin{figure}
\label{fig:tdens}
  \includegraphics[height=.3\textheight]{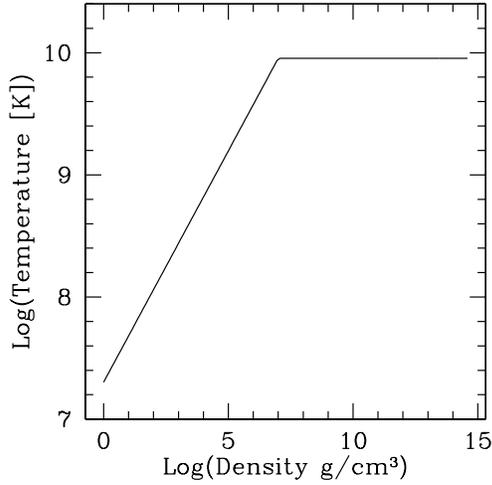}
  \caption{The initial temperatures for the nuclear reaction
    network. The initial temperature depends on the local density of
    the neutron-star crust. Above $10^7$g/cm$^3$ the 
    temperature is just the core temperature and below this density the
     temperature is an interpolation of the surface temperature
    and the central temperature. This is for a central temperature of
    $10^{10}$K.}
\end{figure}

The neutron star cools for one year starting from a central
temperature of $10^{10}$K and after one year the central temperature is  
$9.5\times10^{8}$K. During this first year of the neutron star the
energy budget is assumed to be dominated by the modified Urca process.
Using the Urca process the cooling steps are determined\cite{bhwdns}: 
\begin{equation}
\Delta t = 1{\rm yr} \ T^{-6}_9(f)\left\{ 1-\left[ \frac{T_9(f)}{T_9(i)}
  \right]^6 \right\} \label{urca}
\end{equation} 
where $T_9(f)$ is the temperature of the outer core ($T_{\rm c}$) in
units of $10^9$K.  The temperature used for each of the time steps
depends on the density.  For the densities above $10^7$g/cm$^3$, the
temperature is given by $T=T_{\rm c}$; whereas for the densities below
$10^7$g/cm$^3$ the temperature is interpolated between the surface
($T_{\rm s}$) and the core temperatures by
\begin{eqnarray}
\log(T) &  = &
\left[\frac{\log(T_{\rm c})-\log(T_{\rm s})}{7}\right]\log(\rho) \label{interp}
\\*
 & & {} +\log(T_{\rm s}) \nonumber 
\end{eqnarray}
where the surface temperature is given by \cite{bhwdns}
\begin{equation}
T_{\rm s} = (10\times T_{\rm c})^{2/3}
\end{equation}
where both temperatures are given in Kelvin.  The relationship, in
equation~{\ref{interp}}, between
the initial temperature of  various densities is shown in
Fig.~{\ref{fig:tdens}}, for a central temperature of $10^{10}$K of the
neutron star.

\begin{figure}
\label{fig:cool}
  \includegraphics[height=.3\textheight]{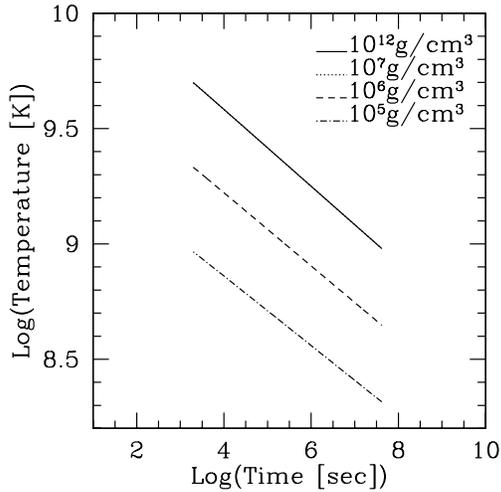}
  \caption{Cooling curve for the neutron star starting with a central
    temperature of from 5$\times 10^9$K and cooling for one year. The
    starting temperature depends on the density of the neutron
    star. For densities above 10$^7$g/cm$^3$ the cooling curves follow
    the same relationship and thus for the curves for
    10$^{12}$g/cm$^3$ and 10$^{7}$g/cm$^3$ overlap. As the densities
    decrease (the depth is closer to the neutron star surface), the
    initial temperature also decreases.}
\end{figure}

The cooling curves of the first year of the neutron star are displayed
in figure {\ref{fig:cool}}. These cooling curves are representative of
the various densities at which the nuclear reactions were calculated.
The temperature for each of these curves is the temperature at the
specified density, as opposed to the central temperature of the
neutron star. For the densities above $10^7$g/cm$^3$ the cooling
curves have the same relationship and thus overlap in the figure. 

In order to calculate the nuclear mass fractions the system is initially set
to nuclear statistical equilibrium with an initial core temperature of
$T_{\rm c}= 10^{10}$K. As the star cools the mass fractions
of the isotopes at the various densities are calculated. Pressure
arguments are used in order to calculate a minimum initial abundance
required for the isotope to be optically thick on the neutron star
surface.  The total
column density of a particular isotope depends on its partial pressure. The
total pressure at a specific density is: 
\begin{equation}
\label{eq:press}
P=\frac{1.2435\times 10^{15}}{\mu_e^{4/3}}\rho^{4/3}, 
\end{equation}
assuming that the star is in the regime of relativistic electrons and
the dominant species, $^{56}$Ni, is fully ionized ($\mu_e$ = 2). By
dividing the partial pressure of a particular isotope by the surface
gravity of the neutron star its column density can be found.  The
surface gravity ($g_{\rm ns}$) used in this analysis assumes a neutron star with a
mass of 1.4M$_\odot$ and a radius of 10 km: $g_{\rm ns}=2.43\times
10^{14}$cm/s$^2$. The minimum mass fraction abundance for which enough
of the isotope
would have risen to the surface for the isotope to have a surface
density of 1~g/cm$^{2}$ and be optically thick is found by dividing the
minimum surface density by the column density of the layer.

\section{Initial Results}

\begin{figure}
\label{fig:surf12}
  \includegraphics[height=.3\textheight]{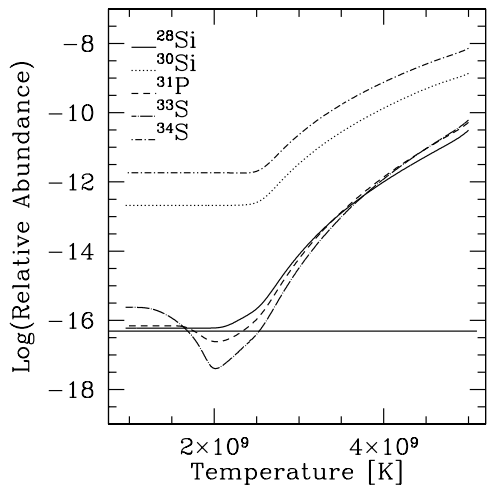}
  \caption{Mass fractions for the density of $10^{12}\;$g/cm$^3$, this
    has a corresponding pressure of  4.9$\times 10^{30}$ dyne and a
    column density of $2.0 \times 10^{16}$g/cm$^2$. The horizontal
    line indicates the minimum mass fraction abundance required in
    order for there to be  a surface density of an isotope of
    1g/cm$^2$. The mass fractions included in this plot are the
    lightest isotopes for which the isotope would be optically thick
    on the surface.}
\end{figure}

We have calculated the mass fractions for two densities within the
neutron-star crust, $10^{12}\;$g/cm$^3$ and $10^7\;$g/cm$^3$. At each of
these densities the neutron star cool for a year starting at a central
temperature of $10^{10}$K. For the case of the higher density,
$10^{12}\;$g/cm$^3$, the corresponding pressure is $4.9\times
10^{30}$dyne, which results in a column density of $2.0 \times
10^{16}$g/cm$^2$. For any mass fraction abundance above $4.9\times
10^{-17}$ enough of the isotope will rise the the neutron star surface
to have at least a surface density of 1g/cm$^2$. We find that at a
density of $10^{12}\;$g/cm$^3$ the lightest elements to rise to the
surface and be optically thick are $^{28}$Si, $^{30}$Si, $^{31}$P, $^{33}$S, and
$^{34}$S. This is shown in figure
\ref{fig:surf12}, where the horizontal line indicates the minimum mass
fraction abundance required for the surface density to be 1g/cm$^2$.

\begin{figure}
\label{fig:surf7}
  \includegraphics[height=.3\textheight]{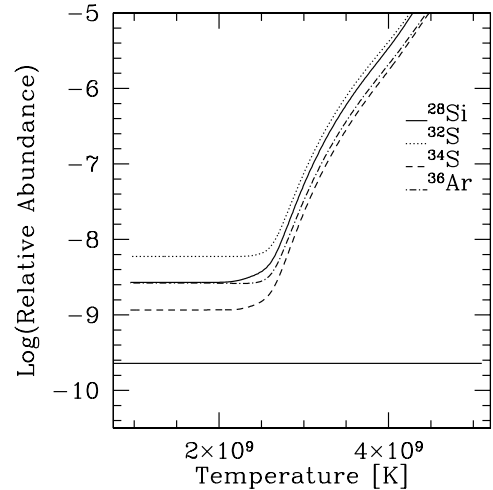}
  \caption{Lightest isotopes for which the mas fraction abundance
    would be great enough that the isotope will be optically thick on
    the neutron star surface. These are the mass fractions for the
    density of $10^7 \;$g/cm$^3$. The horizontal line indicates the
    minimum abundance required for an isotope to have a surface density of
    1g/cm$^2$. The corresponding pressure and column density for this
  neutron star density are $1.1 \times 10^{24}\;$dyne and 4.4$\times
  10^{9}$g/cm$^2$, respectively.}
\end{figure}

Likewise, for the case of the neutron star density of $10^7$g/cm$^3$
we find the corresponding pressure to be $1.1 \times 10^{24}\;$dyne
and the column density to be $4.4\times 10^{9}$g/cm$^2$. A minimum
mass fraction abundance of $2.3 \times 10^{-10}$ at a density of
$10^7$g/cm$^3$ is required for an isotope to have a surface density of
1g/cm$^2$ and be optically thick. The isotopes $^{28}$Si, $^{32}$S,
$^{34}$S, and $^{36}$Ar are the lightest isotopes found to
have large enough abundances to be optically thick after rising to the
surface. Figure \ref{fig:surf7} displays the mass fraction abundances
with a horizontal line indication the minimum abundance of the mass
fractions required.

\section{Conclusions and Future Work}

We have examined the mass fractions that will result in an optically
thick surface layer 
by looking at two cases: a density of $10^{12}\;$g/cm$^3$ and
another of $10^{7}\;$g/cm$^3$. The mass fractions at these densities
are calculated using a 489 isotope reaction network which burns up to
technetium. The neutron star is cooled for a year, starting with a
temperature of $10^{10}$K, assuming the modified Urca process
dominates. Upon calculating the mass fractions the minimum mass
fraction abundance required for enough of an isotope to form an
optically thick layer such that
it has a surface density of 1g/cm$^2$ is calculated. From the two
cases we found that the deeper density resulted in a larger variety of lighter
isotopes can form an optically thick layer. We found $^{28}$Si, $^{30}$Si,
$^{31}$P, $^{32}$S, $^{33}$S, $^{34}$S, and $^{36}$Ar rise to the surface. Future work on
this includes exploring more densities, comparing to analytic
calculations and estimating the timescale for the light impurities to
float to the surface.

\begin{theacknowledgments}
We would like to thank Ed Brown for helpful discussions during the
conference. The Natural Sciences and Engineering Research Council of
Canada, Canadian Foundation for Innovation and the British Columbia
Knowledge Development Fund supported this work. This research has made
use of NASA's Astrophysics Data System Bibliographic Services. 
\end{theacknowledgments}


\bibliographystyle{aipprocl} 


\bibliography{crust}

\begin{thebibliography}{1}
\providecommand{\enquote}[1]{``#1''}
\expandafter\ifx\csname url\endcsname\relax
  \def\url#1{\texttt{#1}}\fi
\expandafter\ifx\csname urlprefix\endcsname\relax\def\urlprefix{URL }\fi

\bibitem{Hern84b}
L.~Hernquist, and J.~H. Applegate, \emph{apj} \textbf{287}, 244 (1984).

\bibitem{heylmnras01}
J.~S. {Heyl}, and L.~{Hernquist}, \emph{M.N.R.A.S.} \textbf{324}, 292--304
  (2001).

\bibitem{timmesapjs99}
F.~X. {Timmes}, \emph{Ap.J. (Suppl)} \textbf{124}, 241--263 (1999).

\bibitem{bhwdns}
S.~L. {Shapiro}, and S.~A. {Teukolsky}, \emph{{Black Holes, White Dwarfs and
  Neutron Stars: The Physics of Compact Objects}}, Black Holes, White Dwarfs
  and Neutron Stars: The Physics of Compact Objects, by Stuart L.~Shapiro, Saul
  A.~Teukolsky, pp.~672.~ISBN 0-471-87316-0.~Wiley-VCH , June 1986., 1986.

\end{thebibliography}


\end{document}